\documentclass[12pt,twoside,fleqn]{article}
\usepackage{epsfig}
\usepackage{multirow}
\usepackage{exscale}
\def\lsim{\raise0.3ex\hbox{$<$\kern-0.75em\raise-1.1ex\hbox{$\sim$}}}
\def\gsim{\raise0.3ex\hbox{$>$\kern-0.75em\raise-1.1ex\hbox{$\sim$}}}

\newcommand{\beqn}{\begin{equation}}
\newcommand{\eqn}{\end{equation}}
\newcommand{\bqa}{\begin{eqnarray}}
\newcommand{\eqa}{\end{eqnarray}}
\newcommand{\bqas}{\begin{eqnarray*}}
\newcommand{\eqas}{\end{eqnarray*}}
\newcommand{\bdm}{\begin{displaymath}}
\newcommand{\edm}{\end{displaymath}}

\begin{document}
\thispagestyle{empty}
\mbox{} \hfill BI-TP 2002/13
\begin{center}
{{\large \bf Heavy Quark Anti-Quark Free Energy and the \\[2mm]
Renormalized Polyakov Loop}
} \\
\vspace*{1.0cm}
{\large O. Kaczmarek, F. Karsch, P. Petreczky and F. Zantow
}

\vspace*{1.0cm}
{\normalsize
$\mbox{}$ {Fakult\"at f\"ur Physik, Universit\"at Bielefeld,
D-33615 Bielefeld, Germany}
}
\end{center}
\vspace*{1.0cm}
\centerline{\large ABSTRACT}

\baselineskip 20pt

\noindent
We calculate the colour averaged and colour singlet
free energies of static quark anti-quark sources placed in 
a thermal gluonic heat bath. We discuss the renormalization of 
these free energies using the short distance properties 
of the zero temperature heavy quark potential. This leads to 
the definition of the renormalized Polyakov loop as an order
parameter for the deconfinement phase transition of the SU(3)
gauge theory which is well behaved in the continuum limit.

\vfill
\noindent
\eject
\baselineskip 15pt

\noindent

\section{Introduction}

Universal properties of the finite temperature phase transition in 
non-abelian $SU(N)$
gauge theories generally are discussed in terms of the Polyakov
loop, which is an order parameter for the confinement-deconfinement
transition. In a somewhat loosely defined terminology the Polyakov loop, 
$L$, is said to be related to the free energy of a heavy, static quark put 
as a test charge into a thermal medium, $L\sim \exp(-F_q/T)$. 
However, it is well known that
the Polyakov loop calculated on the lattice is a priori not properly
renormalized. Taking the continuum limit at fixed temperature will lead
to a vanishing Polyakov loop expectation value even in the deconfined
phase as there are divergent self-energy contributions to the free
energy. A proper renormalization of the Polyakov loop thus is needed
in order to relate it to the heavy quark free energy in the continuum
limit which then can also be used to define a proper order parameter
that survives the continuum limit and can, for instance, be used to 
construct effective actions for the confinement-deconfinement
transition \cite{Pisarski}. 

We suggest here that a proper renormalization of Polyakov loop can be 
obtained through the renormalization of the finite temperature
free energy of a static quark anti-quark pair calculated at
short distances. At short distances the quark anti-quark pair interacts 
through the exchange of gluons, which
may be calculated perturbatively. More importantly we expect that this
interaction is essentially temperature independent for separations $r$ 
which are smaller than the average separation between partons in the
thermal medium. At short distances the finite temperature
heavy quark anti-quark free energy thus will be given by the zero
temperature heavy quark potential. This allows to remove divergent
self-energy contributions in the free energy, $F_{\bar{q}q}(r,T)$, 
through a matching of its short distance behaviour to that of the
heavy quark anti-quark potential, $V_{\bar{q}q} (r)$, at zero temperature. 
After having done so also the large distance behaviour of $F_{\bar{q}q}(r,T)$
is fixed. As this, in turn, is related to the Polyakov loop expectation value, 
it allows to define a renormalized Polyakov loop,
{\it i.e.} as usual no additional divergences will
show up in the heavy quark free energy at finite temperature,
once it has been renormalized properly at zero temperature.

It should be obvious that the program outlined above equally well holds
for the heavy quark free energy calculated in QCD, {\it i.e.} in the
presence of light dynamical quarks. It will be of even greater
importance in this case as $F_{\bar{q}q}(r,T)$ will be finite for all
temperatures and its large distance behaviour in the low temperature,
chiral symmetry broken phase, will reflect the temperature 
dependence of the string breaking energy. In fact, the conceptual
approach we are going to present and analyze here has already been 
anticipated in the presentation of heavy quark anti-quark free energies 
given in Ref.~\cite{Peikert}.

This paper is organized as follows: 
In Section 2 we summarize basic definitions for colour averaged and singlet
free energies and outline our approach to renormalize the Polyakov loop
expectation value. In Section 3 we present a detailed
calculation of the heavy quark anti-quark free energy at short distances.
These calculations are performed within the SU(3) gauge theory 
on lattices with temporal extent up to $N_\tau =16$ so that 
small distances in units of the temperature can be resolved, $rT \sim
1/N_\tau$. We also point out the usefulness of the colour singlet free 
energy for the matching to the heavy quark potential at short distances 
and the determination of the asymptotic behaviour of free energies at
large distances. In Section 4 we discuss the properties of the
renormalized Polyakov loop. Section 5 contains our conclusions. 

\section{Heavy quark anti-quark free energy}

The free energy of a pair of static quark anti-quark sources in a 
thermal medium \cite{McLerran}, 
\begin{equation}
{F_{\bar{q}q} (r,T) \over T} =  -\ln\biggl(
\langle {\rm \tilde{Tr}} L_{\vec{x}} {\rm \tilde{Tr}} L^{\dagger}_{\vec{y}} \rangle
\biggr)
\; + \; c(T)
\quad,\quad rT=|\vec{x}-\vec{y}|/ N_\tau  \quad ,
\label{hqfreeenergy}
\end{equation}
is represented in terms of the
Polyakov loop, which on an Euclidean lattice of size $N_\sigma^3 \times
N_\tau$ is defined as a product of gauge field variables 
$U_{x_0,\vec x}$,

\begin{equation}
L_{\vec x} = \prod_{x_0=1}^{N_\tau}  U_{x_0,\vec x} \quad .
\end{equation}
In Eq.~\ref{hqfreeenergy} and in what follows we use the 
notation $\tilde{{\rm Tr}}={1\over 3} {\rm Tr}$.
Eq.~\ref{hqfreeenergy} defines the colour averaged free energy up
to an additive normalization constant,
$c(T)$, which is related 
to the self-energy of the quark and anti-quark sources.
The Polyakov loop expectation value, 
$\langle L\rangle  \equiv \langle 
N_\sigma^{-3} \;\sum_{\vec x}\; {\rm\tilde{Tr}} L_{\vec x}\rangle $,
is closely related to the long
(infinite) distance behaviour of the free energy,
\begin{equation}
F_\infty (T) = \lim_{r\rightarrow \infty} F_{\bar{q}q} (r,T) 
=-T\; \ln |\langle L \rangle |^2 + c(T)\; T
\quad .
\label{finfinity}
\end{equation}
It is obvious that we need to fix the normalization constant $c(T)$
in order to give a physical meaning to this free energy.

In the pure SU(3) gauge theory the expectation value of the Polyakov 
loop, 
provides an order parameter for the deconfinement phase transition;
in the confined phase $\exp(-F_\infty(T)/2T)$ vanishes, while it stays
non-zero in the deconfined phase.
We would like to interpret $F_\infty$ as the change in free energy
due to the presence of two well separated quarks, each of which is
screened by a cloud of gluons. In order to do so we have
to clarify the normalization of 
$F_{\bar{q}q} (r,T)$
which unambiguously fixes the T-dependent additive constant in
Eq.~\ref{hqfreeenergy}. Before doing so it, however, is worthwhile
to discuss in a bit more detail the properties of
$F_{\bar{q}q} (r,T)$. 

In addition to the colour averaged potential defined through 
Eq.~\ref{hqfreeenergy} we introduce the
colour singlet potential \cite{Nadkarni1},
\begin{equation}
{F_1 (r,T) \over T} =  -\ln\biggl(
\langle \tilde {\rm Tr} L_{\vec{x}} L^{\dagger}_{\vec{y}} \rangle
\biggr)
\; + \; c'(T) \quad .
\label{f1}
\end{equation}
As it stands this correlation function is gauge dependent; for its
evaluation we have to fix a gauge. However,
it recently has been shown that a gauge independent definition of the
singlet free energy
can be given \cite{Philipsen} and that this definition
coincides with the definition given by Eq.~\ref{f1} in gauges which
have a positive transfer matrix representation on the lattice.
Calculating $F_1$ from the Polyakov loop correlation function 
given in Eq.~\ref{f1} in, for
instance, the Coulomb gauge thus yields the singlet potential.

The colour averaged as well as the singlet
free energy will approach finite, non-zero values for all temperatures
$T>T_c$ (and diverge below $T_c$). In fact, if the separation between the
quark anti-quark sources gets large 
the relative orientation of the charges in colour space will not
influence the screening of the individual charges. We thus choose
the relative normalization of $F_1$ and $F_{\bar{q}q}$ such 
that they are identical in the limit of large spatial separations.
The colour averaged free energy defined   
in Eq.~\ref{hqfreeenergy} may then be represented by a thermal average 
over free
energies of a $\bar{q}q$-pair in singlet ($F_1$) and octet ($F_8$)
representations, respectively \cite{McLerran,Nadkarni1},
\begin{equation}
\exp(-{F_{\bar{q}q} (r,T) / T}) = 
{1 \over 9} \exp (- {F_1 (r,T) / T}) +
{8 \over 9} \exp (- {F_8 (r,T) / T}) \quad .
\label{f1f8}
\end{equation}
At distances much shorter than the inverse temperature ($rT<<1$) the
dominant scale is set by $r$, the running coupling will be controlled
by this scale, $g(r,T)\simeq g(r,T=0)\equiv g(r)$, and will become small 
for $r << 1/\Lambda_{QCD}$. 
In this limit the singlet and octet 
free energies are dominated by one-gluon exchange and are calculable 
within ordinary zero temperature perturbation theory,
\begin{equation}
F_1 (r,T=0) =  - 8 F_8 (r,T=0) + {\cal O}(g^4) = -{g^2(r) \over 3\pi r} 
\biggl( 1 + {\cal O}(g^2) \biggr) 
\quad , 
\label{f1pert}
\end{equation}
where $F_1(r,T=0) \equiv V_{\bar{q}q}(r)$ is the static, zero
temperature heavy quark potential.
As the singlet potential is attractive and the octet potential is
repulsive at short distances it follows from Eq.~\ref{f1f8} that in this
limit the colour averaged free energy will be dominated by the singlet
contribution. We then may deduce from Eq.~\ref{f1f8} also the 
asymptotic short distance behaviour of $F_{\bar{q}q}$ and $F_1$,
\begin{equation}
\lim_{r \rightarrow 0} (F_{\bar{q}q} (r,T) - F_1 (r,T)) = T\ln 9 \quad
{\rm for ~all}~T.     
\label{favasym}
\end{equation}        

From Eqs.~\ref{f1pert} and \ref{favasym} it follows that $F_{\bar{q}q}$
as well as $F_1$ will show Coulomb-like behaviour at short distances
and up to a normalization constant will approach the zero temperature 
heavy quark potential.
We note, that this power-like short distance behaviour of the
free energies is quite different from the behaviour of
the colour averaged free energy at large distances ($rT>>1$) where
the dominant scale for the running coupling is set by the temperature
($g(r,T)\equiv g(T)$) and high temperature perturbation theory is used
to show that the leading order contribution to $F_{\bar{q}q}$ is given
by two-gluon exchange \cite{McLerran,Nadkarni2}. 
In general the statement is that
\begin{equation}
{\Delta F_{\bar{q}q}(r,T)\over T} \sim 
\biggl( {\Delta F_1(r,T)\over T} \biggr)^2
\quad {\rm for} \quad  rT\; >>\; 1~~,
\label{asymp}
\end{equation}
where $\Delta F_i \equiv F_i(r,T) -F_\infty$ with $i=1,~\bar{q}q$.
In a somewhat loose notation it thus often is argued that
$F_{\bar{q}q}/T \sim (F_1/T)^2$ \cite{McLerran,Nadkarni2} at large distances.
In the spirit of our previous discussion this statement, however, has to 
be formulated a bit more carefully as $F_\infty$ in general will not be
zero. When fixing the overall
normalization of the free energies at short distances one no longer
has the freedom to assume that they approach zero at large distances.
In this limit the colour averaged as well as the singlet
free energy will approach a finite value, $F_\infty (T)$. 
Using this asymptotic value for the heavy quark anti-quark free energy
we now can define the renormalized Polyakov loop,
\begin{equation}
L^{\rm ren} (T) = \exp (-F_\infty (T)/2T) \quad .
\label{Lren}
\end{equation}
It is obvious that
$L^{\rm ren} (T)$ is zero below $T_c$ on a lattice with infinite
spatial extent, as $F_\infty $ will be infinite in the confined
phase. Moreover, as the renormalization performed at short
distances only involves zero temperature physics it does not
influence the non-analytic structure of $F_\infty$ which will
show up in the vicinity of $T_c$, the renormalized Polyakov
loop. $L^{\rm ren}$ thus will also reproduce all the universal
properties extracted in the past from unrenormalized Polyakov loops
\cite{Engels}.

\section{Numerical results for heavy quark free energies}

We have calculated the colour singlet and colour averaged free energy
of a heavy quark anti-quark pair in a $SU(3)$ 
gauge theory on lattices of size $N_\sigma^3\times N_\tau$ with 
$N_\sigma = 32$ and $N_\tau = 4,~8$ and 16. For our numerical
calculations we used a tree-level improved gauge action
which includes the standard Wilson plaquette term and
planar six link loops. This action has previously been used extensively for
the calculation of thermodynamic quantities \cite{tc}. These
calculations also provided detailed information on the non-perturbative
$\beta$-function determined through the string tension $\sigma a^2$.
We use this information as well as the relation $T_c/\sqrt{\sigma}=0.635$
\cite{tc} to fix the temperature scale.
In order to get access also to
the singlet potential we have fixed the Coulomb gauge.

\begin{figure}[t]
\begin{center}
\epsfig{file=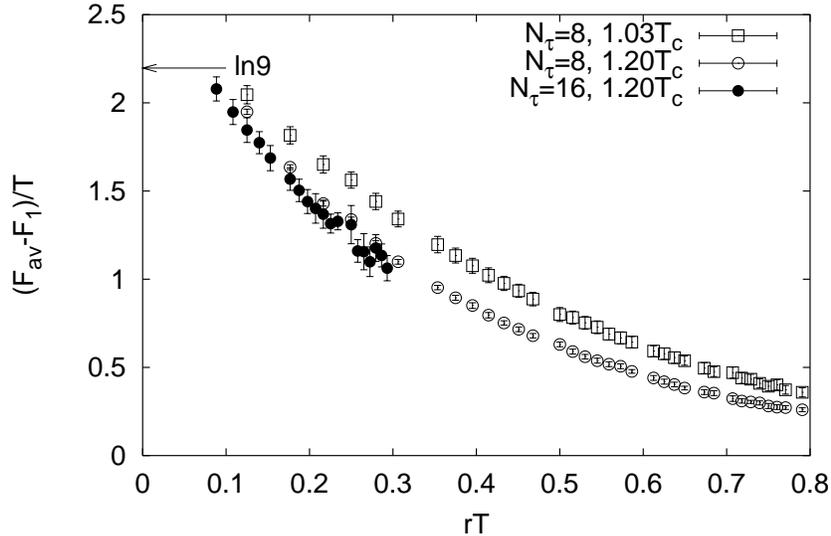,width=110mm}
\end{center}
\caption[]{The difference of the colour averaged and singlet heavy quark 
free energies calculated at two temperatures above $T_c$ calculated on
lattices of size $32^3\times N_\tau$. The free 
energies have been assumed to coincide in the limit of infinite
separation between the quark and anti-quark sources.}
\label{fig:f1mfav}
\end{figure}

Most of our calculations have been performed in the deconfined
phase of the SU(3) gauge theory. The colour averaged and singlet
free energies, calculated according to the definitions given in
Eqs.~\ref{hqfreeenergy} and \ref{f1}, approach a constant value
at infinite distance. As argued above the free energy should not
depend on the relative orientation of the quarks in colour space 
and we thus have fixed the relative normalization of the free energies 
such that they coincide in this limit.
We then can analyze the spatial dependence
of the difference $(F_{\bar{q}q}(r,T) - F_1(r,T))/T$. 
For large distances this difference 
now vanishes by construction and we should find in the short distance
limit, $rT\rightarrow 0$, the relation given in
Eq.~\ref{favasym}. As shown in Figure~\ref{fig:f1mfav}  
the difference indeed approaches the value $\ln 9$ as expected.
This also shows that the colour averaged potential
at short distances indeed is dominated by the singlet contribution.
For $T\; \gsim \; T_c$ and distances $rT \; \lsim\; 0.1$ we find that the 
difference deviates by less than 10\% from the asymptotic value, $\ln 9$.
We also note that the shortest distance that can be resolved on a lattice
with finite temporal extent $N_\tau$ is $rT\; = \; 1/N_\tau$. In order to 
analyze the short distance behaviour of the potential for 
$rT\; \equiv\; 0.1$ it thus was mandatory to use the large temporal lattices
which we have used here for the first time to study the 
finite temperature heavy quark free energies.

\begin{figure}[t]
\begin{center}
\epsfig{file=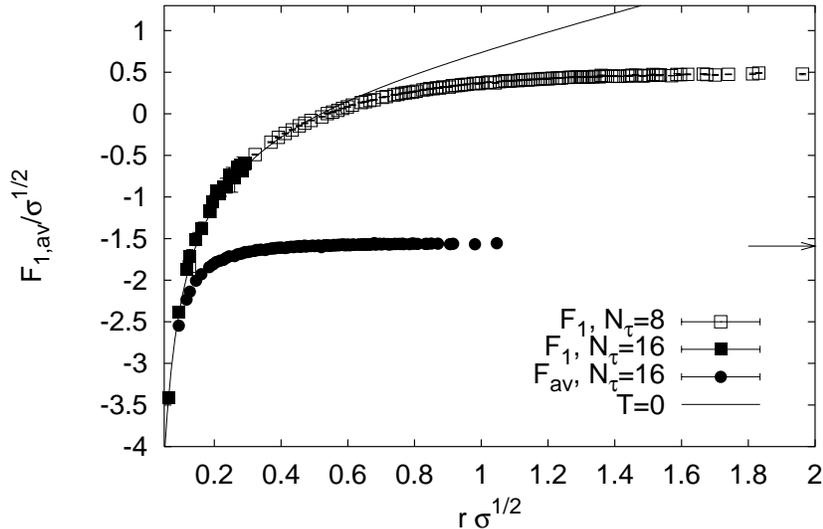,width=110mm}
\end{center}
\caption[]{The shifted colour averaged 
($F_{\rm av}\equiv F_{\bar{q}q} - T\ln 9$) 
and singlet ($F_1$) heavy quark anti-quark free energies calculated at 
$T=1.5\; T_c$ on lattices with temporal extent $N_\tau = 8$ and 16.
The free energies and the spatial separation $r$ have been expressed
in units of the square root of the string tension.
The free energies have been matched to the zero temperature
heavy quark potential (solid line) at the shortest distance
available. The arrow points at the value $(F_\infty-T\ln 9)/\sqrt{\sigma}$,
with $F_\infty$ determined from the colour singlet potential.
} 
\label{fig:f1fav_1.5}
\end{figure}

As discussed in the previous section we expect that at short distances
the colour averaged heavy quark anti-quark free energy is dominated
by the singlet contribution and, moreover, approaches the zero
temperature heavy quark potential, $V_{\bar{q}q}(r)$. In order
to compare our finite temperature calculations of the free energies
with $V_{\bar{q}q}(r)$ we first have to specify the latter. 
Using lattices with small lattice spacing $V_{\bar{q}q}(r)$ has recently 
been calculated for the SU(3) gauge theory for distances larger than 
$0.05$fm and the results have been extrapolated to the continuum
limit \cite{Necco1}. For distances larger than a scale $r_0$,
which is defined through the slope of the heavy quark potential
\cite{Sommer},
\begin{equation}
r_0^2 \;\biggr( {{\rm d} V_{\bar{q}q}(r) \over {\rm d} r}\biggl)_{r=r_0}
= 1.65 \quad ,
\end{equation} 
it is known that $V_{\bar{q}q}$ is well described by the  
simple linear confining potential corrected by a Coulomb-like term
arising from string fluctuations, {\it i.e.}
$V_{\bar{q}q}(r)=\sigma r-{\pi/ 12r}$ for $r>r_0$ with 
\begin{equation}
\sigma r_0^2=1.65-\pi/12 \quad. 
\label{sig2r0}
\end{equation}
For distances smaller than $r_0$ we use a polynomial
interpolation of the lattice data of Ref. \cite{Necco1} normalized
such that the resulting potential smoothly joins the confinement
potential for $r > r_0$, {\it i.e.} we fix the free constant in the
lattice results such that 
$r_0 V_{\bar{q}q}(r=r_0)=1.65-{\pi\over 6} $.
In some cases we also need the zero temperature
potential at distances smaller than $0.1r_0 \simeq 0.05$fm. Here
we use the perturbative 3-loop calculation of the potential
in so-called $qq$ scheme \cite{Necco2} which agrees well with
lattice calculations up to distances $0.25 r_0$ \cite{Necco2}.
For the comparison of free energies with the zero temperature
potential we give them in units of $\sqrt{\sigma}$ which
is straightforward since $r_0$ and $\sqrt{\sigma}$ are related through
Eq.~\ref{sig2r0}. This zero temperature potential is shown in 
Figure~\ref{fig:f1fav_1.5} as a solid line. 

For the matching of free energies to $V_{\bar{q}q}$ 
we also use calculations of $F_1 (r, T)$ and $F_{\bar{q}q}(r,T)$ 
at off-axis separations on the lattice. At short distances these
calculations show violations of rotational symmetry, also known 
from other studies of short distance properties on the lattice.
As the matching is done at short 
distances we thus should try to adjust for these lattice artefacts. 
Following Ref. \cite{Necco1} we replace 
$F_{1,{\bar{q}q}}(r)$ by $F_{1,{\bar{q}q}}(r_I)$ 
where $r_I^{-1}(r)=4 \pi \int_{-\pi}^{\pi}
{d^3 k\over {(2 \pi)}^3 } \exp(i \vec{k}\cdot \vec{r}) D^{(0)}_{00}(k)$,
with $D^{(0)}_{\mu \nu}$ denoting the tree level lattice gluon propagator
evaluated in \cite{Weisz}. We thus 
replace the lattice separation $r$ 
by the separation $r_I$ which corrects for the violation of the 
rotational symmetry in the Coulomb potential calculated on the lattice.

A comparison of the colour averaged free energy ($F_{\bar{q}q}/\sqrt{\sigma}$),
the singlet free energy ($F_1/\sqrt{\sigma}$) and the heavy quark
potential ($V_{\bar{q}q}/\sqrt{\sigma}$) is given in Figure~\ref{fig:f1fav_1.5}.
For clarity we show in this figure the shifted colour averaged
free energy, $F_{\rm av}(r,T) = F_{\bar{q}q}(r,T) - T\ln 9$, which should
coincide with the heavy quark potential at short distances. 
From Figure~\ref{fig:f1fav_1.5} it is evident that indeed the 
shifted colour averaged free energy and the singlet free energy
approach the zero temperature heavy quark potential at short
distances. The good agreement of $F_{\rm av}/\sqrt{\sigma}$ and 
$F_1/\sqrt{\sigma}$ at short distances, furthermore, supports our
assumption that the singlet and colour averaged ($F_{\bar{q}q}/\sqrt{\sigma}$ 
rather than $F_{\rm av}/\sqrt{\sigma}$) free energies approach the same 
constant at large distances. We also note that the color singlet
potential calculated for $N_{\tau}=8$ and $N_{\tau}=16$ agree well
with each other indicating that
residual cutoff effects are small.
Similar results hold for all temperatures examined by us.

\begin{figure}[t]
\begin{center}
\epsfig{file=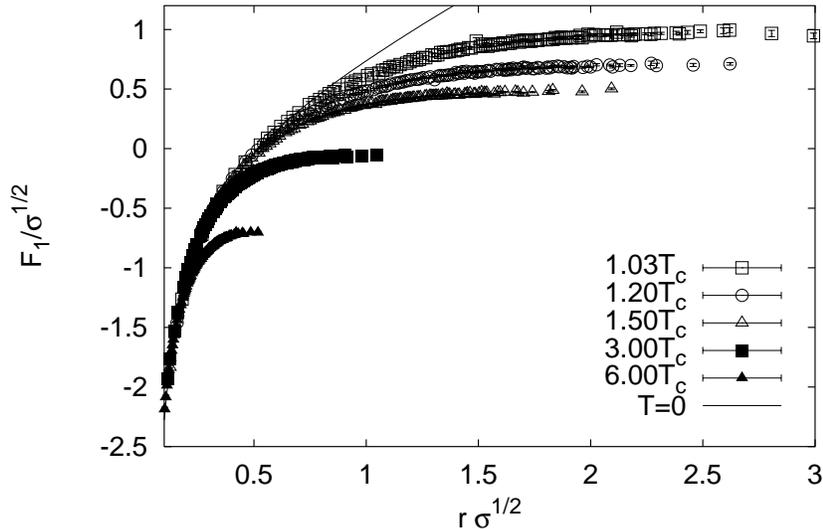,width=110mm}
\end{center}
\caption[]{The colour singlet heavy quark anti-quark free energy ($F_1$)
in units of $\sqrt{\sigma}$ versus $r\sqrt{\sigma}$ calculated on
lattices of size $32^3\times 8$ for several values of the temperature 
above $T_c$. The corresponding gauge couplings are given in Table 1.
The free energies have been matched to the zero temperature
heavy quark potential (solid line) at the shortest distance available.}
\label{fig:norm}
\end{figure}

\section{The renormalized Polyakov loop}

The analysis of the heavy quark anti-quark free energies presented in 
the previous section shows that both can be renormalized by 
matching their short distance behaviour to that of the zero temperature
heavy quark potential. In this way all divergent self-energy
contributions are removed. Moreover, we have noted that
the asymptotic large distance behaviour of $F_1(r,T)$ coincides 
with that of $F_{\bar{q}q}(r,T)$. The former thus equally well determines 
the asymptotic value of the free energy, $F_\infty (T)$.  As it is
obvious from
Figure~\ref{fig:f1fav_1.5} that it is much easier to perform the
matching to the zero temperature potential at short distances using
the colour singlet free energy $F_1(r,T)$ we have used this free energy
rather than $F_{\bar{q}q}(r,T)$ to determine the asymptotic value of
the free energies, $F_\infty$. In particular, this allows to perform
the matching to $V_{\bar{q}q}$ already with finite temperature
free energies calculated on lattices with temporal extent $N_\tau =4$.
Results for the colour singlet free energy calculated at several
values of the temperature above $T_c$ are shown in
Figure~\ref{fig:norm}. From these renormalized free energies
we have determined 
$F_\infty (T)$ at different values of the temperature. Actually, in
order to avoid any fits, we use the value of $F_1(r,T)$ at
maximal on-axis separation of the quark anti-quark sources on a
lattice with spatial extent $N_\sigma$, {\it i.e.} we define 
$F_\infty (T)\equiv F_1(N_\sigma/2,T)$. Systematic errors on the
free energies have been estimated by matching them to $V_{\bar{q}q}$
at distance $rT=1/N_\tau$ as well as $rT=\sqrt{2}/N_\tau$.
Some results are collected in Table~\ref{tab:free}.
\begin{table}[t]
\begin{center}
\begin{tabular}{|c|c|c|c|}
\hline       
$T/T_c$ &$\beta$ &  $F_\infty (T) /T$ & $r_{\rm screen}T$ \\
\hline
1.03 & 4.5592 & 1.42(4) & 0.59 \\
1.20 & 4.6605 & 0.93(4) & 0.58 \\
1.50 & 4.8393 & 0.53(4) & 0.62 \\
3.00 & 5.4261 & -0.03(3) & 0.80 \\
6.00 & 6.0434 & -0.18(1) & 0.98 \\
\hline
\end{tabular}
\end{center}
\caption{Change in free energy due to the presence of a heavy quark
anti-quark pair in a thermal gluonic heat bath. The table gives
results for $F_\infty (T) /T$ obtained in calculations on a $32^3\times 8$ 
lattice and the screening radius defined through Eq.~\ref{rscreen}. 
Errors on $F_\infty (T) /T$ include an estimate of the systematic errors
(see text). 
The temperatures corresponding to the gauge couplings $\beta$
are given in the first two columns.} 
\label{tab:free}
\end{table}

As can be seen in Figure~\ref{fig:norm} the singlet free energy rapidly
changes from the Coulomb like short distance behaviour to a constant
value. This, of course, reflects the exponential screening of the two
static charges. As this change is so rapid, we may use $F_\infty (T) /T$
to define a screening radius $r_{\rm screen}$, which characterizes
the onset of screening, through the relation

\begin{equation}
{V_{\bar{q}q}(r_{\rm screen}) \over T} \equiv  {F_\infty (T) \over T} \quad .
\label{rscreen}
\end{equation}
The values for $r_{\rm screen} / T$ determined in this way are also given
in Table~\ref{tab:free}. As can be seen $r_{\rm screen} \simeq 0.4$~fm
in the vicinity of $T_c\simeq 270$MeV and drops to $0.2$~fm
for $T=3\; T_c$. We do expect that asymptotically the screening radius
will drop like the inverse Debye mass, {\it i.e.}
$r_{\rm screen} \sim 1/g(T)T$.

\begin{figure}[t]
\begin{center}
\epsfig{file=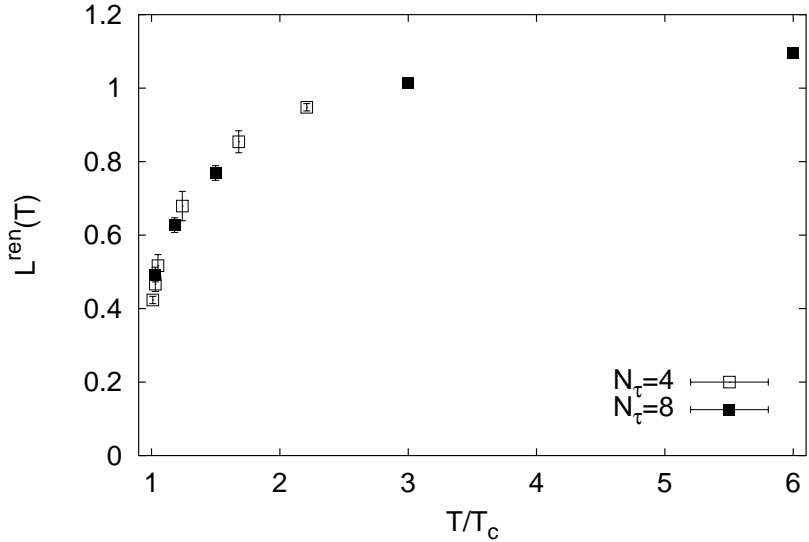,width=110mm}
\end{center}
\vskip -0.3truecm
\caption[]{The renormalized Polyakov loop expectation value 
defined in Eq.~\ref{Lren} determined from the asymptotic behaviour
of colour singlet free energies on lattices of size $32^3\times N_\tau$.}
\label{fig:Lren}
\end{figure}

Using the values for $F_\infty(T)$ we can now construct the renormalized 
Polyakov loop defined in Eq.~\ref{Lren}. Results are shown in
Figure~\ref{fig:Lren}.
The renormalized Polyakov loop is an order parameter for the
deconfinement phase transition, which is well defined also in the
continuum limit. Its magnitude is related to the free energy of 
a heavy quark placed in a thermal gluonic heat bath. It is obvious
from Table~\ref{tab:free} that $F_\infty/T$ becomes negative at large 
temperatures and consequently $L^{\rm ren}$ becomes larger than
unity. We note, however, that $L^{\rm ren}$ is fixed only up to a 
multiplicative renormalization which results from fixing an arbitrary
additive constant in the zero temperature heavy quark potential.

\section{Outlook}

In this paper we have defined the renormalized Polyakov loop by
matching the free energy of a static quark anti-quark pair
at short distances to the zero temperature heavy quark potential.
We have shown that the renormalized Polyakov loop can be determined
from the large distance behaviour of the colour averaged as well
as the colour singlet free energy of the $\bar{q}q$-pair. The 
approach has been used here to study the heavy quark free energy
of the SU(3) gauge theory. It, however, generalizes without any
difficulties to the case of QCD.  

In the temperature regime analyzed by us, $T/T_c \le 6$, the 
renormalized Polyakov loop is a monotonically rising function.
It becomes larger than unity for $T/T_c \simeq 2.5$. In the future
it will be interesting to analyze the asymptotic behaviour of
the Polyakov loop at even larger temperatures and determine its
infinite temperature limit.  In this limit $L^{\rm ren}$ is expected
to approach a constant. Figure~\ref{fig:Lren} suggests that this
constant is close to unity. A more detailed analysis of the large
temperature behaviour of $L^{\rm ren}$ would also 
allow to make contact with perturbative calculations, which suggest that 
the asymptotic value may be approached from above \cite{self}.

Finally we note that our normalization of the heavy quark anti-quark free
energies at short distances also opens the possibility for a new look at the
heavy quark potential at finite temperature.
Using the thermodynamic relations between entropy, energy and free energy, 
$S=-\partial F / \partial T$, $U=-T^2\partial(F/T)/ \partial T$, it is 
evident from Figure 3 that $S_{\bar{q}q}(r,T)$ vanishes at short
distances while it clearly is positive at large distances. It therefore
will add a positive contribution to the total energy, 
$U_{\bar{q}q}(r,T) = F_{\bar{q}q}(r,T) + T S_{\bar{q}q}(r,T)$,
which becomes 
larger with increasing separation $r$.
This shows that the change in energy due to the presence of a heavy
quark anti-quark pair in a thermal bath is quite complex. In particular,
its $r$-dependence is not only given by $F_{\bar{q}q}(r,T)$.   
This indicates that it may be misleading to use $F_1(r,T)$
in potential models for the study of heavy quark bound states.

\section*{Acknowledgements}
We thank Edwin Laermann for very helpful discussions.
This work has been supported by the DFG under grant FOR 339/1-2. It
was in part based on the MILC collaboration's public lattice
gauge code. See http://physics.utah.edu/$\sim$detar/milc.html

\end{document}